\documentclass[twocolumn,showpacs,preprintnumbers,amsmath,amssymb,pre]{revtex4}

\usepackage{graphicx}
\usepackage{dcolumn}
\usepackage{bm}

\begin{document}


\title{Synchronization properties of self-sustained mechanical
oscillators}

\author{Sebasti\'an I. Arroyo$^1$ and Dami\'an H. Zanette$^{1,2}$}
\affiliation{$^1$Instituto Balseiro and Centro At\'omico Bariloche, \\
8400 San Carlos de Bariloche, R\'{\i}o Negro, Argentina
\\
$^2$Consejo Nacional de Investigaciones Cient\'{\i}ficas y
T\'ecnicas, Argentina}

\date{\today}

\begin{abstract}
We study, both  analytically and numerically, the dynamics of
mechanical oscillators kept in motion by a feedback force, which is
generated electronically from a signal produced by the oscillators
themselves. This kind of self-sustained systems may become standard
in the design of frequency-control devices at microscopic scales.
Our analysis is thus focused on their synchronization properties
under the action of external forces, and on the joint dynamics of
two to many coupled oscillators. Existence and stability of
synchronized motion are assessed in terms of the mechanical
properties of individual oscillators --namely, their natural
frequencies and damping coefficients-- and synchronization
frequencies are determined. Similarities and differences with
synchronization phenomena in other coupled oscillating systems are
emphasized.
\end{abstract}


 \maketitle

\section{Introduction}

In electronic devices, time keeping and event synchronization rely
upon one or more components able to provide cyclic signals, which
are used as frequency references. Since mid-twentieth century,
quartz crystals were ubiquitously employed in this function and
became standard in the construction of clocks of all kinds. At
micrometric scales and below, however, technical difficulties in the
fabrication and mounting of quartz crystals motivate considering
alternative solutions, preferably based on simpler components.
Micromechanical oscillators --tiny vibrating bars of semiconductor
material, which can be readily integrated into electronic circuits
during manufacturing, and kept in motion by very small electric
fields-- are an attractive possibility \cite{sci,rev}.

\begin{figure}
\includegraphics[width=.6\columnwidth]{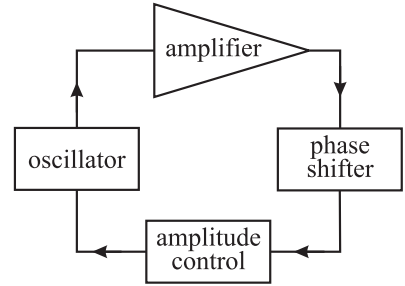}
\caption{Self-sustained mechanical oscillator. The displacement of
the oscillator with respect to its equilibrium position is
electronically read, amplified, and shifted in phase. The signal is
then reinjected as an amplitude-controlled force acting on the
oscillator. Adapted from Ref.~4. } \label{fig1}
\end{figure}

In order to function as a frequency reference, an oscillator must
perform sustained periodic motion at a frequency determined by its
own dynamics (i.e., independent from any external signal). A
feedback mechanism able to produce self-sustained motion in a
mechanical oscillator \cite{Greywall,Yurke} is inspired on the
well-known phenomenon of resonance: under the action of external
periodic forcing, the response of the oscillator, measured by the
amplitude of its motion, is maximal if the external force and the
oscillator's velocity are in-phase or, equivalently, if the
oscillator's displacement from its equilibrium position is a quarter
of a cycle late with respect to the force \cite{reso}. The feedback
self-sustaining mechanism consists in electronically reading the
displacement of an autonomous mechanical oscillator, and advancing
the signal by a quarter of a cycle, namely, by a positive phase
shift $\phi_0 = \pi/2$ (or, equivalently, a delay of $3\pi/2$). In
practice, this shifting in phase can be achieved in a variety of
ways --for instance, using resistive circuits or all-pass filters
\cite{NatCom}. The shifted signal is then reinjected as a force
acting on the oscillator. To avoid the effect of damping --which
would eventually lead the oscillator to rest-- the amplitude of the
force must be controlled externally and, ultimately, maintained by a
battery. Under the action of this conditioned signal, the oscillator
moves with maximal amplitude at a frequency determined by its
internal parameters (and, possibly, by the amplitude of the
self-sustaining feedback force). Figure \ref{fig1} shows a schematic
representation of the self-sustaining circuit.

The oscillator's motion is conveniently described by the Newton
equation for a coordinate $x(t)$, representing the displacement from
equilibrium:
\begin{equation} \label{Newt1}
m \ddot x + \gamma \dot x+ k x + \eta (x) = F_0 \cos (\phi+\phi_0)
+F (t),
\end{equation}
where $m$, $\gamma$, and $k$ are effective values for the mass, the
damping, and the elastic constant. The term $\eta (x)$ stands for
non-elastic forces. The first term in the right-hand side of the
equation represents the self-sustaining force. As discussed above,
its amplitude $F_0$ is an independent parameter, determined by the
feedback mechanism. The self-sustaining force depends on the phase
of the oscillator's motion $\phi$, which is defined on the basis
that, in harmonic oscillations, $x(t)$ is proportional to $\cos \phi
(t)$ (see Sect. \ref{fase}). The phase shift between the force and
the coordinate should ideally be fixed at $\phi_0=\pi/2$ but, in
order to assess the effect of this parameter on the oscillator's
dynamics, it is here allowed a generic value. Note that, as a
function of $\phi$, the form of the self-sustaining force is not
aimed at modeling any specific experimental implementation of the
phase shifting, but rather at representing its effect on the
reinjected conditioned signal. In addition to $\eta (x)$, the force
$F_0 \cos (\phi+\phi_0)$ is also a source of nonlinearity: while its
phase is directly related to that of $x(t)$, its amplitude is
independent of the motion. Finally, $F(t)$ denotes any additional
force that may be acting on the oscillator, ranging from
externally-applied deterministic signals to electronic noise and
thermal fluctuations.

A key technological problem associated with the use of
self-sustained mechanical oscillators in micro-devices is the
instability of their frequency under the effect of noise
\cite{Yurke,noise} and of changes in the amplitude of oscillations
\cite{af,af2}. Coupling several oscillators to obtain a collective,
more robust signal to be used as a frequency reference could be a
plausible solution to this problem. It is therefore of much interest
to study the synchronization properties of a set of oscillators of
this kind interacting with each other.

Our aim in this paper is to characterize the dynamics of
self-sustained mechanical oscillators, as governed by
Eq.~(\ref{Newt1}). In order to focus on the role of the
self-sustaining mechanism, we disregard the nonlinearities
represented by the non-elastic forces $\eta (x)$ --although we
recognize their importance in the functioning of this kind of
oscillators at microscopic scales \cite{NatCom}. After establishing
analytical and numerical procedures to deal with the oscillation
phase $\phi$ as a dynamical variable, we first study the properties
of self-sustained motion of a single oscillator. Then, taking into
account the remark in the previous paragraph, we concentrate on
synchronized dynamics in various situations: a self-sustained
oscillator under the action of a harmonic external force, two
oscillators coupled to each other, and an ensemble of globally
coupled oscillators. Our conclusions emphasize similarities and
differences with collective motion in other kinds of coupled
dynamical systems.

\section{Dynamics of the self-sustained oscillator} \label{SS}

Assuming that non-elastic forces are absent, $\eta (x) \equiv 0$,
and redefining the time unit as the inverse of the natural
oscillation frequency $\Omega_0 =\sqrt{k/m}$ and the coordinate unit
as $F_0/k = F_0/m\Omega_0^2$, Eq.~(\ref{Newt1}) can be recast as two
first-order equations for the coordinate $x(t)$ and its velocity
$v(t)$,
\begin{eqnarray}
\dot x &=& v, \nonumber \\
\dot v &=& - x - \epsilon v + \cos(\phi+\phi_0) +f (t),
\label{Newt2}
\end{eqnarray}
with $\epsilon = \gamma/m\Omega_0$ and $f(t) =F(t)/F_0$. Apart from
the quantities that determine the self-sustaining and external
forces, the only parameter in these equations is the rescaled
damping coefficient $\epsilon$. It is worth pointing out that
$\epsilon$ coincides with the inverse of the oscillator's quality
(or Q-) factor: $\epsilon= Q^{-1}$. We recall that the quality
factor $Q$ is a non-dimensional measure of the resonance bandwidth
relative to the resonance frequency, and characterizes the rate of
energy dissipation as compared with the oscillation period. In
experiments involving micromechanical oscillators \cite{NatCom},
typical values are around  $Q \sim 10^4$.

It is important to realize that Eqs.~(\ref{Newt2}) do not specify a
dynamical system, for the coordinate and the velocity, in the usual
sense. Indeed, as will become clear in the following, the
oscillator's phase $\phi (t)$ cannot be unambiguously defined in
terms of $x(t)$ and $v(t)$ alone --although $\phi (t)$ does
represent an instantaneous property of the motion. To solve the
equations, in any case, an operational definition of the phase
becomes necessary. In the next section we discuss how analytical and
numerical approaches prompt different ways to deal with this
question.

\subsection{Analytical treatment and numerical evaluation of the phase}
\label{fase}

In a typical experiment, even with a high quality factor, a
micromechanical oscillator will be found in its long-time asymptotic
dynamical regime after at most a few seconds. With $Q = 10^4$ and a
frequency $\Omega = 5 \times 10^4$ Hz \cite{NatCom}, for instance,
any transient regime associated with dissipative effects fades out
with a characteristic relaxation time $\tau \sim Q\Omega^{-1} =0.2$
s. If the asymptotic motion is harmonic, to all practical purposes,
the oscillation phase is thus well defined when the oscillator's
output signal is fed into the phase shifter to construct the
self-sustaining feedback force (see Fig.~\ref{fig1}). Tuning the
phase shifter allows the experimenter to apply a prescribed shift to
the oscillation with no need to measure the phase itself.

On the other hand, both in the analytical and in the numerical
treatment of the equations of motion (\ref{Newt2}), it is necessary
to specify the value of the phase at each time, in order to be able
to calculate the instantaneous self-sustaining force. The
determination of that value must also work during transients or in
non-harmonic motion, whose occurrence cannot be discarded a priori
when solving the equations.

Analytically, a convenient way to deal with the dependence of the
self-sustaining force on the oscillation phase is to introduce
$\phi$ itself as one of the variables of the problem. This is
achieved by replacing the original coordinate-velocity variables
$(x,v)$ by a set of phase-amplitude variables $(\phi,A)$ through a
canonical-like transformation \cite{Talman},
\begin{equation}
x(t) = A(t) \cos \phi(t) , \ \ \ \ \ v(t) = -\nu A(t) \sin \phi(t) .
\end{equation}
The arbitrary constant $\nu$, which can adopt any real value,
parameterizes the variable transformation. As we explain below, it
can be chosen in such a way as to make certain solutions of the
equations of motion attain a simple mathematical form. The change of
variables transforms Eqs.~(\ref{Newt2}) into
\begin{eqnarray}
\dot A \cos \phi &=& A (\dot \phi-\nu) \sin \phi  ,\nonumber \\
\dot A \sin \phi &=& A(\nu^{-1}-\dot \phi) \cos \phi -\epsilon A
\sin \phi \nonumber \\  &&- \nu^{-1}  \cos(\phi+\phi_0) -\nu^{-1}f .
\label{Newt3}
\end{eqnarray}
This formulation has the advantage that the phase is defined for any
kind of motion, not only for harmonic oscillations, as $\phi(t) = -
\arctan [v(t)/ \nu x(t)]$. On the other hand, it turns out to depend
on the specific choice of the parameter $\nu$. As a function of the
coordinate and the velocity, as advanced above, the oscillator's
phase is therefore not unambiguously defined. Focusing however on
harmonic motion --which is characterized by constant amplitude $A$
and constant frequency $\dot \phi$-- the first of Eqs.~(\ref{Newt3})
makes it clear that the solution will take a particularly simple
form if $\nu$ is chosen to coincide with the oscillation frequency.
In fact, for $\nu=\dot \phi$ and $\dot A = 0$, that equation is
satisfied automatically, and the problem reduces to solve the second
equation. In some cases --for instance, in synchronized motion under
the action of an external harmonic force (see Sect. \ref{Sync1})--
we know in advance the oscillation frequency, and can therefore
conveniently fix $\nu$ before finding the solution. When, on the
other hand, the frequency is part of the solution itself --as is the
case for an autonomous self-sustained oscillator (Sect.~\ref{SSO}),
or for two mutually coupled oscillators (Sect.~\ref{2OS})-- $\nu$
can be considered as an additional unknown of the problem, and
obtained together with the solutions to the equations of motion.

In the numerical integration of Newton equations (\ref{Newt2}), in
turn, there are no reasons to assume that the frequency of harmonic
solutions is known a priori. Consequently, the phase must be
evaluated from the numerical solution itself, as it is progressively
obtained, without resorting to a specific change to phase-amplitude
variables. The standard method for assigning an instantaneous phase
to the signal $x(t)$, through the construction of its analytical
imaginary part using the Hilbert transform \cite{Hilbert}, is here
ineffectual, as it requires the whole (past and future) signal to be
available at each time where $\phi(t)$ is calculated. We have
instead implemented a numerical algorithm that estimates the
instantaneous phase in terms of the coordinate along the numerical
integration, as follows.

\begin{figure}
\includegraphics[width=.8\columnwidth]{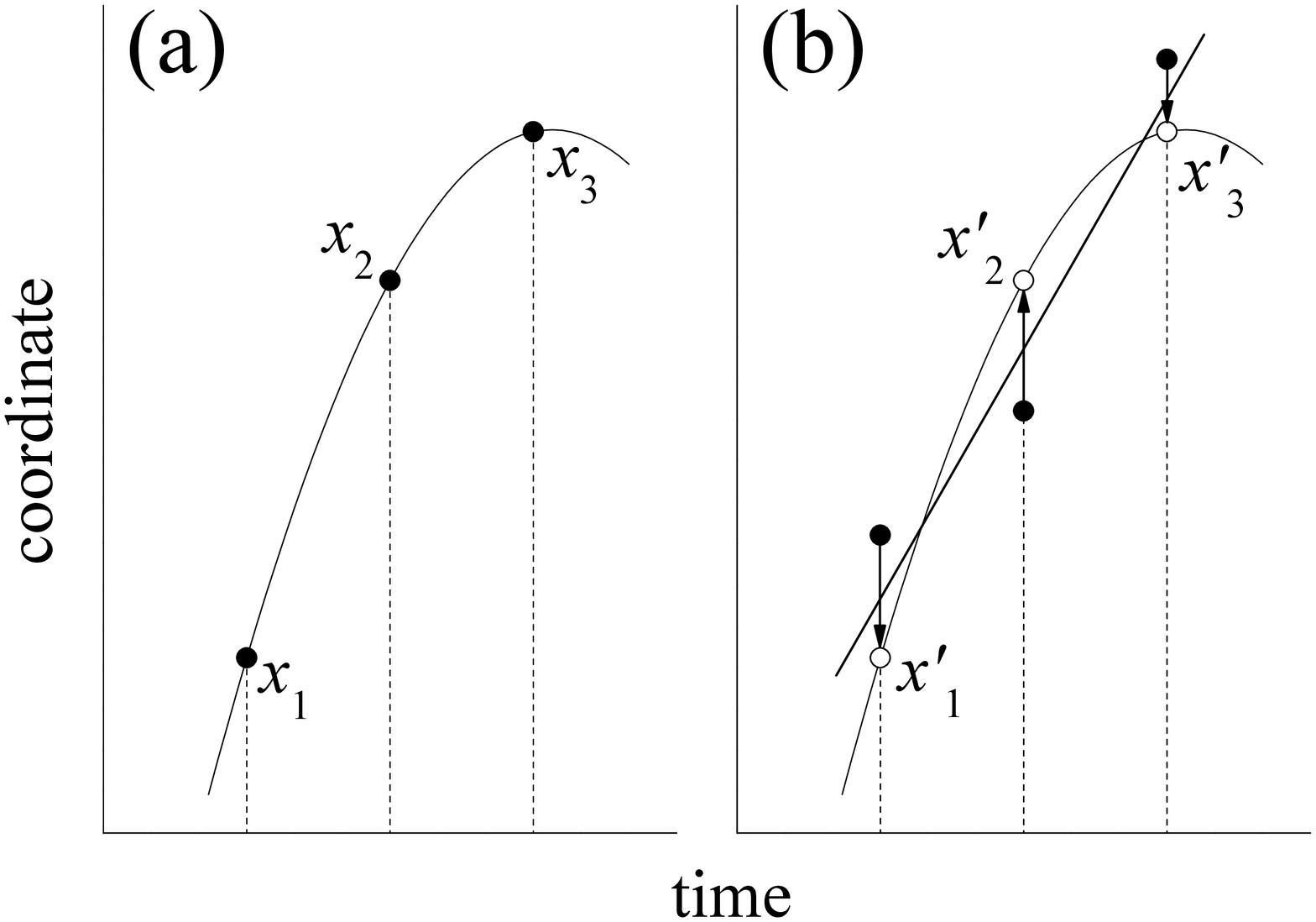}
\caption{Schematic representation of the numerical evaluation of the
oscillation phase. In (a), the coordinates satisfy $|x_1+x_3|\le 2
|x_2|$ and a cosine function exactly fits the three points. In (b),
$|x_1+x_3 | >2 |x_2|$ and the original points (full dots) must be
replaced by auxiliary points (primed coordinates, empty dots),
obtained by reflection of the original points with respect to the
abscissa of their least-square linear fitting (straight line).}
\label{fig2}
\end{figure}

Let $x(t-2h) \equiv x_1$, $x(t-h) \equiv x_2$ and $x(t) \equiv x_3$
be three successive values of the coordinate in the numerical
solution with integration step $h$, and define $t_1\equiv t-2h$,
$t_2\equiv t-h$, and $t_3\equiv t$. Under the condition discussed
below, it is possible to find constants $a$, $w$, and $u$ such that
$x_i = a \cos (w t_i+ u)$ for $i=1,2,3$. Obtaining these constants,
the phase at time $t$, $\phi (t) = w t_3+ u$, is given by
\begin{eqnarray}
\cos \phi (t) &=& \frac{x_3}{2x_2} \sqrt{\frac{4
x_2^2-(x_1+x_3)^2}{x_2^2-x_1x_3}}, \nonumber \\
\sin \phi(t) &=& \frac{2 x_2^2 - x_3 (x_1 + x_3)}{2 x_2
\sqrt{x_2^2-x_1x_3}}. \label{fase3p}
\end{eqnarray}
This solution, however, is well defined only if $|x_1+x_3|\le
2|x_2|$. It is not difficult to realize that this condition is
equivalent to requiring that the three points $x_1$, $x_2$, and
$x_3$ define a curve with the same convexity as the fitting cosine
function. In the case that $|x_1+x_3|>2|x_2|$, we define three
auxiliary points,
\begin{equation}
x'_1= 2 \bar x -x_3,  \ \ \ x'_2= 2\bar x -x_2, \ \ \ x'_3 = 2 \bar
x -x_1,
\end{equation}
with $\bar x =(x_1+x_2+x_3)/3$, which satisfy $|x'_1+x'_3|\le
2|x'_2|$ and can therefore be fitted by a cosine. The auxiliary
points $x'_i$ are reflections of the original points $x_i$ with
respect to the ordinate of their least-square linear fitting, as
illustrated in Fig.~\ref{fig2}b. Both sets of points, $x'_i$ and
$x_i$, have therefore identical linear trends. Consequently, the
value of $\phi (t)$ calculated from Eq.~(\ref{fase3p}) using now the
auxiliary points, is still a satisfactory evaluation of the phase
associated to the points $x_1$, $x_2$, and $x_3$.

Note that our numerical definition of the phase, given by
Eqs.~(\ref{fase3p}), is independent of the integration step $h$. In
fact, it provides a value for $\phi(t)$ for any trajectory
successively passing by the coordinates $x_1$, $x_2$, and $x_3$,
with the only condition that they are equally spaced in time. Since
in the numerical integration of the equations of motion we need to
define the phase at each step, we identify those coordinates with
consecutive values of $x(t)$ along the calculation.

Once the phase $\phi (t)$ at a given integration step has been
evaluated, it is used to calculate the self-sustaining force $ \cos
(\phi+\phi_0)$ and, thus, the numerical increment of the velocity
$v(t)$. Since the evaluation of the phase requires knowing the
coordinate at three successive steps, in our numerical calculations
--which were performed using a second-order Runge-Kutta algorithm--
the self-sustaining force was switched on after the first few
integration steps had elapsed. This procedure had no significant
effect on the subsequent dynamics.

It must be borne in mind that, in general, the analytical definition
of $\phi( t )$ using the phase-amplitude variables, and its
numerical evaluation in terms of three successive values of the
coordinate, are equivalent only when the motion is a harmonic
oscillation. Since this kind of motion is not guaranteed a priori,
analytical and numerical results must be carefully contrasted with
each other when assessing the dynamics of the self-sustained
oscillator.

\subsection{Self-sustained oscillations} \label{SSO}

When no external forces act on the oscillator ($f= 0$), its motion
is controlled by the interplay between its mechanical properties and
the self-sustaining force. Assuming that the long-time asymptotic
motion is a harmonic oscillation of constant amplitude $A$ and
frequency $\dot \phi = \Omega$ --which needs not to coincide with
the natural frequency $\Omega_0 \equiv 1$-- Eqs.~(\ref{Newt3}) yield
\begin{equation} \label{AW-SSO}
A= \frac{ \sin \phi_0}{\epsilon \Omega}, \ \ \ \ \ \Omega=
\sqrt{1+\left( \frac{\epsilon}{2 \tan \phi_0} \right)^2} -
\frac{\epsilon}{2 \tan \phi_0}.
\end{equation}
These solutions were obtained by fixing $\nu  \equiv \Omega$ and
separating, in the second of  Eqs.~(\ref{Newt3}), terms proportional
to $\cos \Omega t$ and $\sin \Omega t$.

\begin{figure}
\includegraphics[width=.7\columnwidth]{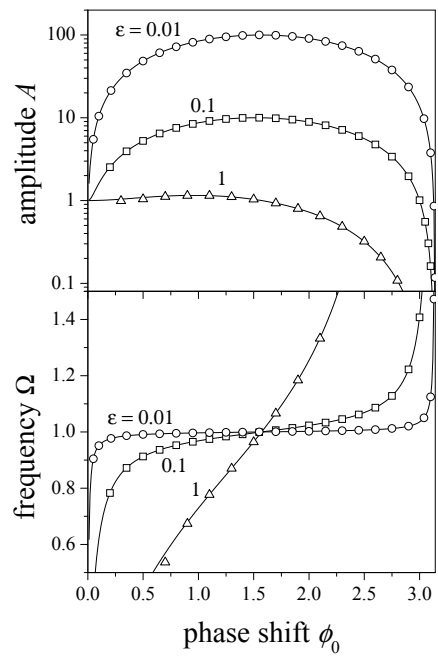}
\caption{Amplitude and frequency of self-sustained harmonic
oscillations as functions of the phase shift of the self-sustaining
force, for three values of the damping $\epsilon$.  The natural
frequency of the oscillator is $\Omega_0=1$. Curves: Analytical
solution to Eqs.~(\ref{Newt3}), as given by Eqs.~(\ref{AW-SSO}).
Symbols: Numerical solution to Eqs.~(\ref{Newt2}), using the
three-point evaluation of the oscillation phase, as described in
Sect.~\ref{fase}.} \label{fig3}
\end{figure}

Figure \ref{fig3} shows the amplitude and frequency of
self-sustained harmonic oscillations as functions of the phase shift
$\phi_0$ of the self-sustaining force, for three values of the
damping $\epsilon$. As advanced in the Introduction, we find that,
for small damping (high quality factor), the amplitude is maximal
when the phase shift is around $\pi/2$. For $\phi_0=\pi/2$, $\Omega$
coincides with the oscillator's natural frequency, $\Omega_0=1$, and
the interval where $\Omega \approx \Omega_0$ grows as the damping
decreases.

Symbols in Fig.~\ref{fig3} represent results obtained from the
numerical integration of the Newton equations (\ref{Newt2}), using
the three-point method described in Sect.~\ref{fase} for the
evaluation of the oscillation phase. In the numerical solutions, the
amplitude and the oscillation frequency were measured by recording
the coordinate and the time at the integration steps where the
velocity changes its sign --from positive to negative, i.e.~at the
coordinate maxima-- and averaging the results over several hundred
oscillation cycles after a sufficiently long transient interval. The
agreement with the analytical solution is excellent, which validates
the assumption of harmonic oscillations.

\subsection{Synchronization with harmonic external forcing} \label{Sync1}

Turning the attention to the simplest situation where the
synchronization properties of self-sustained oscillators are to be
assessed, let us consider a single oscillator subject to the action
of an external force with harmonic time dependence, $f(t)= f_1 \cos
\omega t$. As is the case for a broad class of oscillating systems
\cite{synchro1,synchro2}, the self-sustained oscillator is able to
synchronize with the external force and perform harmonic motion with
frequency $\omega$, provided  that $\omega$ and the self-sustained
frequency $\Omega$, given by the second of Eqs.~(\ref{AW-SSO}), are
close enough to each other.

Looking for harmonic solutions of frequency $\omega$,
Eqs.~(\ref{Newt3})  yield the equation
\begin{equation} \label{complex1}
A\left( \omega^2 -1 + {\rm i}\epsilon \omega \right) =-e^{-{\rm i}
\phi_0} -f_1 e^{-{\rm i} \psi},
\end{equation}
for the amplitude $A$ and phase shift $\psi$ of the coordinate,
$x(t)=A \cos (\omega t-\psi)$. Solutions to this equation exist when
$\omega$ lies in an interval around $\Omega$, whose width grows with
both the damping coefficient $\epsilon$ and the amplitude $f_1$ of
the external force. For $f_1<1$, the interval of existence of
synchronized harmonic solutions is determined by the two frequencies
\begin{equation}
\omega_{\displaystyle_{2}^{1}}=\frac{\epsilon}{ 2\alpha_\pm}
+\sqrt{1+\left(\frac{\epsilon}{ 2\alpha_\pm}\right)^2},
\end{equation}
with
\begin{equation}
\alpha_\pm =\frac{-\sin \phi_0 \cos \phi_0 \pm
f_1\sqrt{1-f_1^2}}{\cos^2 \phi_0 -f_1^2}.
\end{equation}
For $f_1\to 0$, both $\omega_1$ and $\omega_2$ tend to $\Omega$, and
the width of the interval of existence vanishes. For $f_1<\sin
\phi_0 $, the solution exists if $\omega_1 \le \omega \le \omega_2$.
For $f_1 \ge \sin \phi_0$, on the other hand, the solution exist for
all $\omega \le \omega_2$ if $\phi_0 <\pi/2$, and for all $\omega
\ge \omega_1$ if $\phi_0
>\pi/2$. In the limiting case where $\phi_0 = \pi/2$, the interval
of existence is bounded by
\begin{equation} \label{Spi/2}
\omega_{\displaystyle_{2}^{1}} = \sqrt{1+\frac{\epsilon^2 f_1^2}{4
(1-f_1^2)}} \mp \frac{\epsilon f_1}{2 \sqrt{1-f_1^2}}.
\end{equation}
Meanwhile, when the amplitude of the external force is larger than
that of the self-sustaining force, $f_1>1$, synchronized solutions
exist for any frequency $\omega$.

\begin{figure}
\includegraphics[width=.9\columnwidth]{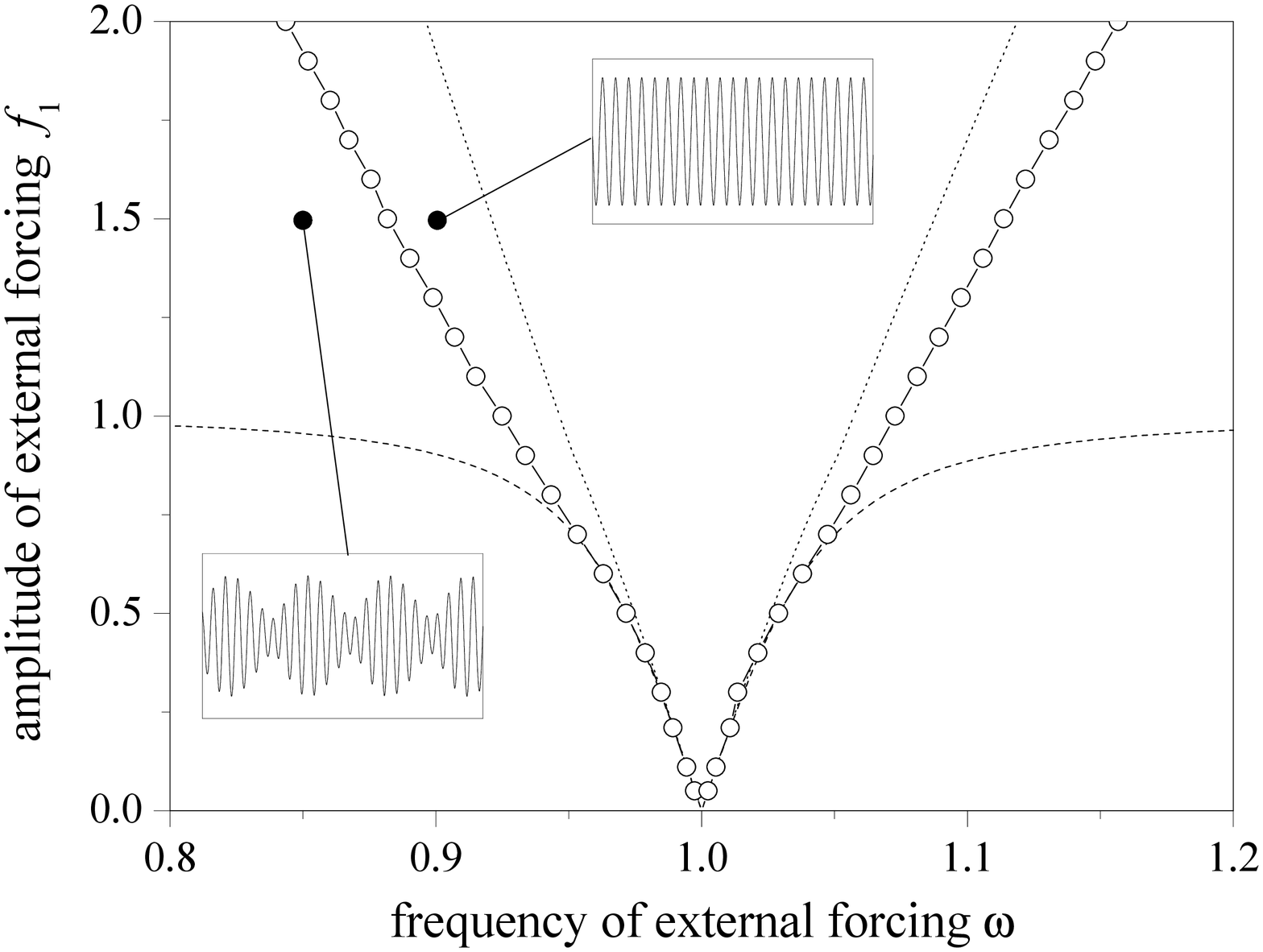}
\caption{Existence and stability of synchronized motion under the
action of an external harmonic force of amplitude $f_1$ and
frequency $\omega$, for a self-sustaining force with phase shift
$\phi_0=\pi/2$. The self-sustained frequency is $\Omega=1$, and
$\epsilon=0.1$. Dashed lines and empty dots delimitate,
respectively, the regions of (analytical) existence and (numerical)
stability. Dotted lines correspond to the analytical approximation
to the stability region given by Eq.~(\ref{Floq}). The insets show
the coordinate $x(t)$ for two parameter sets (full dots), inside and
outside the stability region.} \label{fig4}
\end{figure}

Dashed lines in Fig.~\ref{fig4} delimitate the region in the
parameter plane $(\omega,f_1)$ where synchronized harmonic solutions
exist, for a phase shift $\phi_0=\pi/2$ and $\epsilon=0.1$, as given
by Eq.~(\ref{Spi/2}). Numerical integration of Eqs.~(\ref{Newt2}),
however, shows that inside this ``existence tongue'' the motion is
not always a harmonic oscillation of frequency $\omega$. In other
words, synchronized solutions are actually observed within a
subregion of the tongue only, where they are stable. The boundaries
of the ``stability tongue'' are shown by full dots in
Fig.~\ref{fig4}. They have been found numerically, by dichotomic
search of synchronized solutions for selected values of the force
amplitude $f_1$, until a precision $\delta \omega = 5 \times
10^{-4}$ was reached in the frequency axis. Motion was considered to
be synchronized with the external force when its frequency,
calculated from the average period over $10^3$ oscillation cycles,
differed from $\omega$ by less than $10^{-4}$.

Analytically, the stability of synchronized motion could be
determined by means of Floquet theory \cite{Floquet}, by linearizing
around the time-dependent harmonic solutions. For our two-variable
non-autonomous system, however, the theory is not able to provide an
explicit condition for stability. An approximate criterion can
nevertheless be obtained by replacing all time-periodic terms in the
linearized problem by their respective averages. The resulting
stability condition is
\begin{equation} \label{Floq}
\epsilon+\frac{f_1 \sin \psi}{\omega A} >0,
\end{equation}
where $A$ and $\psi$  are given by Eq.~(\ref{complex1}). Dotted
lines in Fig.~\ref{fig4} show the result of this approximation.
While its description of the stability region is not quantitatively
good, it correctly reproduces the functional trend with the
amplitude of the external force.

The insets in Fig.~\ref{fig4} illustrate the time dependence of the
coordinate  $x(t)$ inside and outside the synchronization tongue,
over a span of $150$ time units. They correspond to an external
force of amplitude $f_1=1.5$, and frequencies $\omega=0.9$ and
$\omega=0.85$, respectively. While, inside the tongue, $x(t)$
maintains a constant oscillation amplitude, the non-synchronized
signal displays beats, resulting from a combination of the frequency
of the external force and the oscillator's self-sustained frequency.

\begin{figure}
\includegraphics[width=.7\columnwidth]{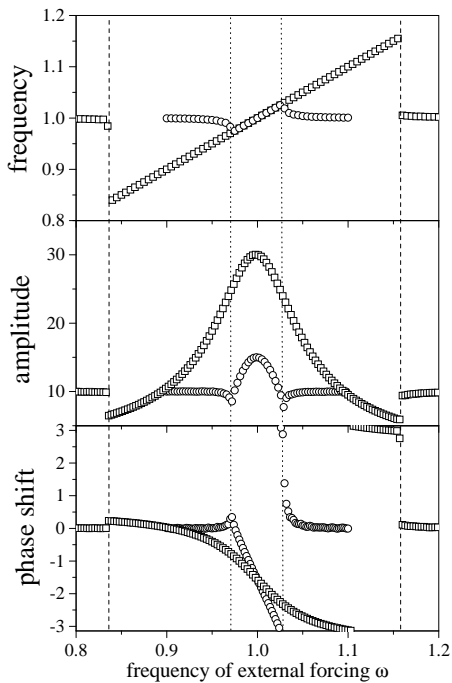}
\caption{Oscillation frequency, amplitude, and phase shift of an
externally forced,  self-sustained oscillator ($\Omega=1,
\epsilon=0.1$), as functions of the frequency $\omega$ of the
external force, for two values of its amplitude, $f_1=0.5$ (circles)
and $2$ (squares). Dotted and dashed vertical lines indicate the
respective boundaries of the synchronization region.} \label{fig5}
\end{figure}

Figure \ref{fig5} shows numerical results for the oscillation
frequency, amplitude, and phase shift for an oscillator of
self-sustained frequency $\Omega=1$ ($\phi_0=\pi/2$) and
$\epsilon=0.1$, as functions of the frequency $\omega$ of the
external force, for two values of its amplitude. Circles and squares
correspond, respectively, to $f_1=0.5$ and $2$. Vertical dotted and
dashed lines indicate the respective boundaries of the
synchronization region. Results where obtained as explained in
Sect.~\ref{SSO} for Fig.~\ref{fig3}, averaging over a large number
of oscillation cycles. Outside the synchronization region, where the
averages are performed over the beating signal, the average
oscillation frequency coincides with $\Omega$ and the amplitude and
phase shift have constant values. For the synchronized signal, on
the other hand, the frequency equals $\omega$, the amplitude attains
a maximum at $\omega=\Omega$, and the phase varies approximately
between $\psi=0$ and $\pi$. Note that, at the amplitude maximum,
$\psi=\pi/2$. The external force and the self-sustaining force are
therefore in-phase at this point, and the amplitude is given by the
sum of the values determined by the two forces,
$A_{\max}=(1+f_1)/\epsilon$. Inside the synchronization range, the
numerical results for $A$ and $\psi$ are in excellent agreement with
the analytical solutions to Eq.~(\ref{complex1}) which, for clarity,
are not shown in the figure.

\section{Mutual synchronization of self-sustained oscillators}

We study now the collective dynamics of a population of mutually
coupled self-sustained mechanical oscillators. From the experimental
viewpoint, a straightforward way to make the oscillators interact
with each other is to substitute the individual self-sustaining
force by a linear combination of the feedback signals of all
oscillators --built up, for instance, by means of a resistive
circuit. To represent this situation, Eqs.~(\ref{Newt2}) are
replaced by a set of equations of motion for each oscillator,
\begin{eqnarray}
\dot x_i &=& v_i, \nonumber \\
\mu_i \dot v_i &=& - \kappa_i x_i - \epsilon_i v_i + \sum_{j=1}^N
f_{ij} \cos(\phi_j+\phi_{0j}) , \label{NewtN}
\end{eqnarray}
for $i=1,\dots ,N$, where $\mu_i=m_i/m_0$,
$\kappa_i=k_i/k_0=k_i/m_0\Omega_0^2$ are, respectively, the
effective mass and elastic constant of oscillator $i$ divided by
reference quantities $m_0$ and $k_0$, used to fix time and
coordinate units (see Sect. \ref{SS}). The normalized damping
coefficient is $\epsilon_i = \gamma_i/m_0\Omega_0$. The coefficient
$f_{ij}$ weights the contribution of the self-sustaining force of
oscillator $j$ to the coupling signal applied to oscillator $i$, and
$\phi_{0j}$ is the corresponding phase shift. The size of the
population is $N$.

The profusion of free coefficients in Eqs.~(\ref{NewtN}) calls for
some simplifying assumptions, both on the diversity of individual
dynamical parameters and on the coupling force. In the following
sections, therefore, we focus on the special situation where all
self-sustaining forces have the same weight in the interaction, and
all oscillators are feeded the same coupling force: $f_{ij}=N^{-1}$
for all $i$, $j$. The dependence of these coefficients on the number
of oscillators $N$ warrants that the coupling force is comparable
between populations of different sizes. Moreover, we assume that the
phase shift is the same for all oscillators, $\phi_{0i} = \phi_0$
for all $i$, and usually consider the maximal-response value
$\phi_0=\pi/2$. The parameters $\mu_i$, $\kappa_i$, and $\epsilon_i$
are, in principle, left to vary freely. In our numerical
simulations, however, we fix $\kappa_i=1$ for all $i$, choosing to
control the diversity of the individual natural frequencies,
$\Omega_i = \sqrt{\kappa_i /\mu_i}$, by means of the mass ratios
$\mu_i$ only.

\subsection{Two-oscillator synchronization} \label{2OS}

The joint dynamics of two coupled oscillators serves as an
illustrative intermediate case between a single oscillator subject
to external forcing and an ensemble of interacting oscillators. We
take the parameters of oscillator $1$ as reference values for
defining time and coordinate units, so that $\mu_1=\kappa_1=1$. Its
natural frequency is, therefore, $\Omega_1=1$.

In synchronized motion, the two oscillators are expected to perform
harmonic oscillations with a common frequency $\Omega_s$ and a
certain phase shift between each other: $x_1(t)= A_1 \cos ( \Omega_s
t -\psi )$, $x_2 (t) =A_2 \cos \Omega_s t$. Using this Ansatz in the
phase-amplitude representation, we get the equations
\begin{eqnarray}
A_1\left( \Omega_s^2 -1 + {\rm i}\epsilon_1 \Omega_s \right) &=& -
e^{-{\rm i} \phi_0} \frac{ 1+e^{ -{\rm
i} \psi}}{2},\nonumber \\
A_2\left( \mu_2 \Omega_s^2 -\kappa_2 + {\rm i}\epsilon_2 \Omega_s
\right) &=&  -   e^{-{\rm i} \phi_0} \frac{ 1+e^{ {\rm i} \psi}}{2},
\label{complex2}
\end{eqnarray}
to be solved for $\Omega_s$, $A_1$, $A_2$, and $\psi$. Compare these
equations with Eq.~(\ref{complex1}).

\begin{figure}
\includegraphics[width=.9\columnwidth]{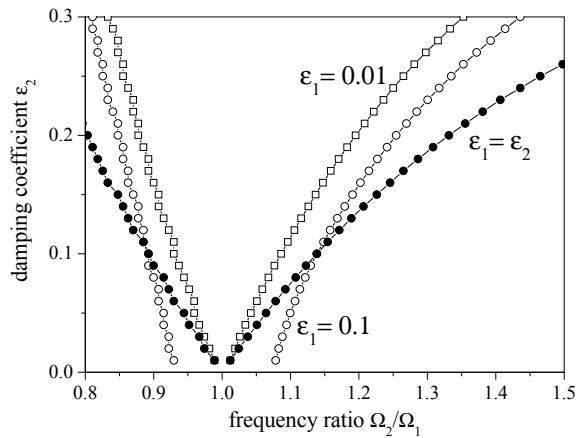}
\caption{Stability regions for synchronized motion of two coupled
oscillators with natural frequencies $\Omega_1=1$ and $\Omega_2$
($\phi_0=\pi/2$). Empty and full symbols correspond, respectively,
to the case of fixed damping coefficient for oscillator $1$, and  to
equal damping for both oscillators.} \label{fig6}
\end{figure}

Solutions to Eqs.~(\ref{complex2}) exist for arbitrary values of all
the involved parameters. Numerical resolution of  the respective
Newton equations, however, reveals that synchronization is observed
only when the self-sustained frequencies of the two oscillators are
sufficiently close to each other. Otherwise, synchronized motion is
unstable. The synchronization range turns out to depend on the
damping coefficients $\epsilon_i$, becoming wider as the damping
increases. Full dots in Fig.~\ref{fig6} show the boundary of the
stability tongue for the case $\epsilon_1=\epsilon_2$, in the plane
whose coordinates are the ratio of the oscillators' natural
frequencies, $\Omega_2/\Omega_1$ (which, for $\phi_0=\pi/2$,
coincide with the self-sustained frequencies), and the damping
coefficient $\epsilon_2$. Empty symbols correspond to the case where
$\epsilon_1$ is fixed and $\epsilon_2$ varies. Results for
$\epsilon_1 = 10^{-3}$, not shown in the plot, are practically
coincident with those for $\epsilon_1 = 0.01$, which thus constitute
a good representation of the limit of very small $\epsilon_1$. For
$\epsilon_1=0.1$, on the other hand, the tongue has an appreciable
size even for very small $\epsilon_2$, indicating that the width of
the synchronization range is controlled by the larger damping
coefficient.

For $\phi_0=\pi/2$, the solution to Eqs.~(\ref{complex2}) for the
synchronization frequency is
\begin{equation} \label{Omega_s}
\Omega_s = \sqrt{\frac{\kappa_2+\epsilon_2/\epsilon_1}
{\mu_2+\epsilon_2/\epsilon_1}}.
\end{equation}
Note that $\Omega_s$ always lies in the interval between
$\Omega_1=1$ and $\Omega_2=\sqrt{\kappa_2 / \mu_2}$ --either for
$\Omega_1 <\Omega_2$ or vice versa. For $\epsilon_2\ll \epsilon_1$
we have $\Omega_s\approx \Omega_2$, while for $\epsilon_2\gg
\epsilon_1$ we have $\Omega_s\approx \Omega_1$. Therefore, the
common frequency of the synchronized oscillators approaches the
natural frequency of the oscillator with the smaller damping
coefficient. The oscillator with the larger quality factor thus
drives synchronized motion.

\begin{figure}
\includegraphics[width=.7\columnwidth]{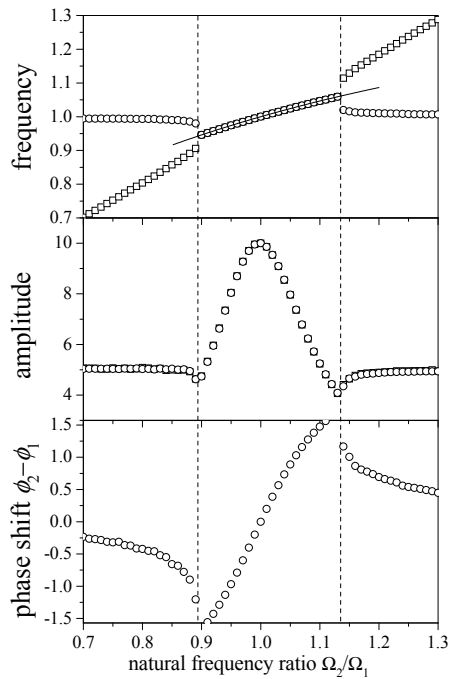}
\caption{Oscillation frequency (upper panel), amplitude (central
panel), and phase shift (lower panel) of two coupled self-sustained
oscillators with $\epsilon_1=\epsilon_2=0.1$ and $\phi_0=\pi/2$, as
functions of the natural frequency ratio $\Omega_2/\Omega_1$.
Vertical dashed lines indicate the boundaries of the synchronization
range. In the two uppermost panels, circles and squares correspond
to oscillators $1$ and $2$, respectively. The curve in the upper
panel represents the analytical result for the synchronization
frequency.} \label{fig7}
\end{figure}

The upper panel in Fig.~\ref{fig7} shows, as symbols, the numerical
evaluation of the oscillation frequencies of two coupled
self-sustained oscillators with $\phi_0=\pi/2$ and
$\epsilon_1=\epsilon_2=0.1$, as functions of the ratio of their
natural frequencies, $\Omega_2/\Omega_1$. They have been calculated
from the numerical solution to Newton equations, as explained in
previous sections. Much as in the case of a single oscillator
subject to an external force (see upper panel of Fig.~\ref{fig5}),
the frequencies change abruptly as the boundary of the
synchronization range is traversed. The curve stands for the
analytical prediction for the synchronization frequency, given by
Eq.~(\ref{Omega_s}).

The central and lower panels in Fig.~\ref{fig7} display numerical
results for the oscillation amplitude and phase shift between the
two oscillators. Since, except for their natural frequencies, the
two oscillators are identical and the coupling force acting over
them is the same, their amplitudes are only slightly different
outside the synchronization range and are exactly coincident inside.
For $\Omega_2=\Omega_1$ the amplitudes reach their maximum,
$A_{\max} = \epsilon^{-1}$. At that point, the two oscillators are
in-phase, $\phi_2-\phi_1=\psi=0$, and their respective contributions
to the coupling force have always the same sign and magnitude. At
the boundaries of the synchronization range, on the other hand, the
phase shift attains values around $\pm \pi/2$. Note, finally, that
for $\Omega_2 < \Omega_1$ and $\Omega_2 > \Omega_1$ we have,
respectively, $\phi_2 - \phi_1 <0$ and $\phi_2 - \phi_1 > 0$. Thus,
irrespectively of whether they are synchronized or not, the
oscillator with the lower natural frequency is always retarded with
respect to its partner.

\subsection{Collective synchronization of oscillator ensembles}

The results obtained in the previous sections suggest that, in a
population of coupled self-sustained oscillators, synchronized
motion would be observed if their individual natural frequencies,
$\Omega_i = \sqrt{\kappa_i/\mu_i}$, are sufficiently close to each
other. It is under this condition that oscillators could become
mutually entrained, so as to perform coherent collective dynamics.
The dispersion of natural frequencies should therefore control the
capability of the ensemble to display synchronization. In a large
population, in any case, the tendency to entrainment of oscillators
with similar frequencies is expected to compete with the disrupting
effect of non-synchronized oscillators, whose incoherent signal
influences the whole system through the coupling force. We show in
this section that this competition gives rise to complex collective
behavior, including partial synchronization in the form clustering.

The only kind of collective motion that can be dealt with with our
analytical tools corresponds to the case of full synchronization of
the whole ensemble, where all oscillators move with the same
frequency $\Omega_s$, but are generally out-of-phase with respect to
each other. Writing the coordinate of each oscillator as $x_i(t)=A_i
\cos(\Omega_s t - \psi_i) $, the phase-amplitude representation of
Newton equations (\ref{NewtN}) yields
\begin{equation}  \label{complex3}
A_i (\mu_i \Omega_s^2-\kappa_i +{\rm i} \epsilon_i \Omega_s) e^{
{\rm i} \psi_i } =
 -\frac{
  e^{-{\rm i} \phi_0 }}{N} \sum_{j=0}^N
e^{{\rm i} \psi_j },
\end{equation}
for $i=1,\dots,N$. The complex factor $z= N^{-1} \sum_j \exp({\rm i}
\psi_j )$ in the right-hand side provides a collective
characterization of the distribution of relative phases in the
oscillator ensemble. Note that its modulus $\rho = |z|$ coincides
with the order parameter used in Kuramoto's theory for phase
oscillator synchronization to quantitatively assess the
synchronization transition \cite{Kuramoto,synchro2}. When phases are
homogeneously distributed over $(0,2\pi)$, we have $\rho \sim
N^{-1/2}$, and $\rho$ grows approaching unity as phases accumulate
toward each other.

\begin{figure}
\includegraphics[width=.9\columnwidth]{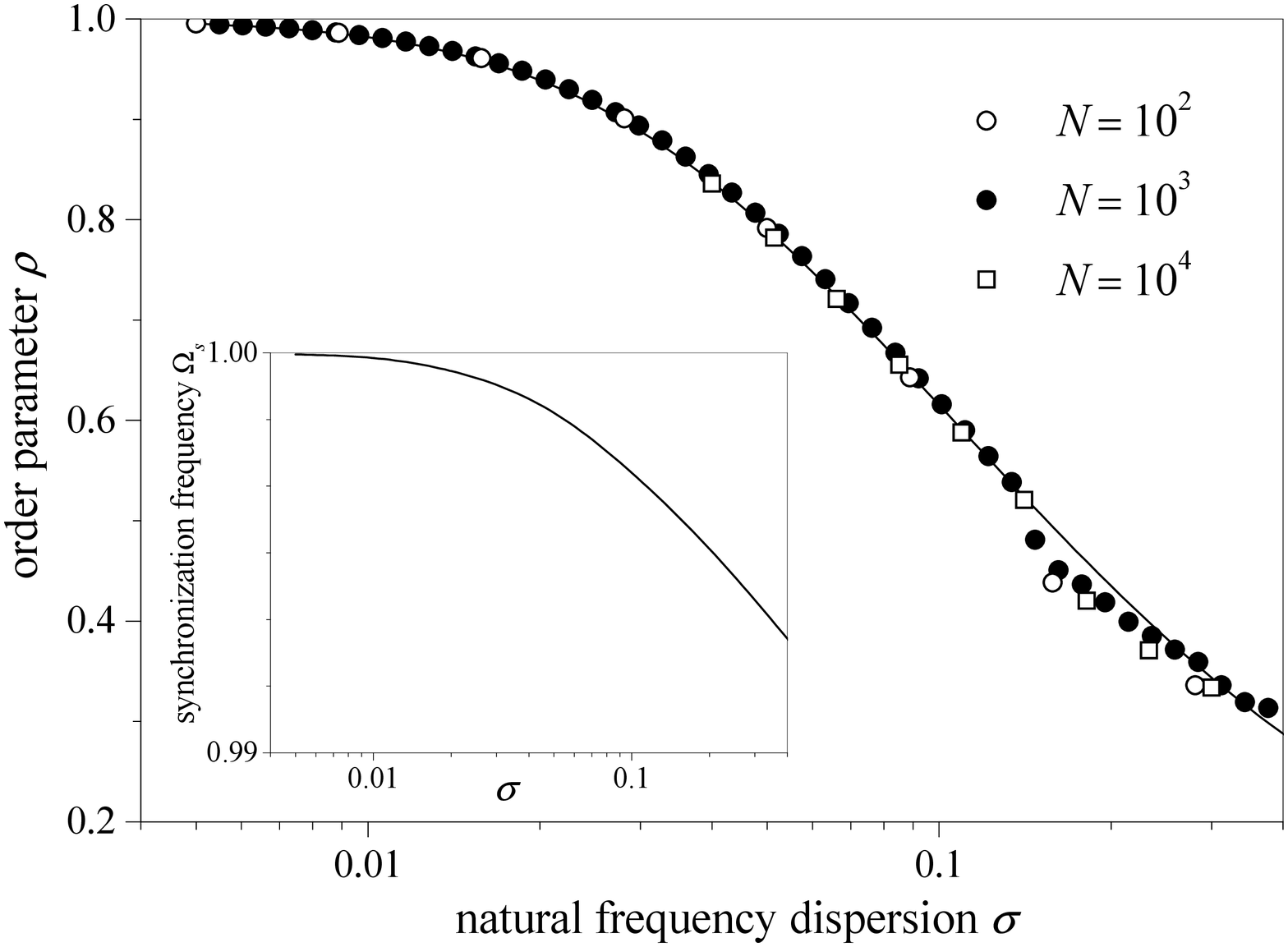}
\caption{The order parameter $\rho$ as a function of the mean square
dispersion of natural frequencies, for an ensemble of size $N$.
Natural frequencies are extracted from a Gaussian distribution with
mean value $\bar \Omega=1$. All oscillators have damping coefficient
$\epsilon_i=0.1$. Dots stand for numerical results for three values
of $N$, and the curve is the analytical result for $ N \to \infty$,
obtained from Eqs.~(\ref{orderpar}) in the continuous limit. The
corresponding result for the synchronization frequency $\Omega_s$ is
shown in the inset.} \label{fig8}
\end{figure}

Equations (\ref{complex3}) make it possible to show that, for
$\phi_0=\pi/2$, the synchronization frequency $\Omega_s$ and the
order parameter $\rho$ satisfy
\begin{eqnarray}
0&=&\sum_{i=1}^N \frac{\mu_i \Omega_s^2  -\kappa_i}{\sqrt{(\mu_i
\Omega_s^2  -\kappa_i)^2 +\epsilon_i^2 \Omega_s^2}} , \nonumber \\
\rho&=&\sum_{i=1}^N \frac{\epsilon_i \Omega_s}{N\sqrt{(\mu_i
\Omega_s^2 -\kappa_i)^2 +\epsilon_i^2 \Omega_s^2}} .\label{orderpar}
\end{eqnarray}
Figure \ref{fig8} and its inset represent, as curves, the order
parameter and the synchronization frequency obtained from these
equations for an ensemble where natural frequencies are drawn from a
Gaussian distribution with mean value $\bar \Omega=1$ and mean
square dispersion $\sigma$, in the limit $N\to \infty$. We have
taken $\kappa_i=1$ for all $i$, so that each natural frequency fixes
univocally the mass ratio $\mu_i$. The damping coefficients are
equal for all oscillators, $\epsilon_i=0.1$ for all $i$. When the
frequency dispersion is small we have $\rho \approx 1$, indicating
that the phase shift between oscillators in the synchronized state
is small as well. As $\sigma$ grows, $\rho$ decreases monotonically,
and phases become less similar to each other. The inset, in turn,
shows that the synchronization frequency changes only slightly from
$\bar \Omega$, decreasing by less that 1\% over two orders of
magnitude of variation in $\sigma$.

Symbols in the main plot of Fig.~\ref{fig8} correspond to the
evaluation of the order parameter from the numerical solution of
Newton equations for three population sizes, with natural
frequencies drawn at random from the same distribution as above. For
dispersions below $\sigma \approx 0.13$, the numerical results for
different sizes are in excellent agreement between themselves and
with the analytical prediction. For $\sigma \gtrsim 0.13$, on the
other hand, numerical and analytical results depart from each other.
Inspection of the ensemble for those values of $\sigma$ shows that
oscillators are in fact not synchronized. Not all of them oscillate
with the same frequency and, consequently, the coherent motion of
phases breaks down. This explains that the order parameter
calculated from the numerical solution of the equations of motion
lies generally below the analytical prediction, which assumes full
synchronization of the ensemble.

Numerical results show that, when full synchronization is not
observed, a part of the ensemble splits into several internally
synchronized clusters. The oscillation frequency of the members of
each cluster --determined, as in previous sections, from the average
period between successive maxima in their coordinates-- is the same,
while the frequency differs between clusters. This state of partial
synchronization is reminiscent of the clustering regime of coupled
chaotic elements, which has been profusely reported for both
continuous-time dynamical systems for maps
\cite{synchro2,clus1a,clus1b,clus2a,clus2b,clus2c}. In our case,
clustering seems to be induced by stochastic fluctuations in the
distribution of natural frequencies: clusters tend to form where the
randomly chosen frequencies become accumulated by chance. Clustering
in ensembles of coupled dynamical elements is a highly degenerate
regime, where the collective state of the system strongly depends on
the specific choice of the individual parameters --in the present
situation, the natural frequencies-- and initial conditions.
Necessarily, therefore, the study of this regime is restricted to
numerical analysis, and to a semi-quantitative illustration of its
main features. While the results presented below pertain to a
specific realization of the oscillator ensemble, they are
demonstrative of the generic behavior within the clustering regime.

We consider an ensemble of $N=100$ coupled oscillators, all with the
same damping coefficient $\epsilon_i=0.1$. In order to be able to do
a detailed comparison of individual dynamics for different values of
the mean square dispersion of natural frequencies, each natural
frequency is defined as a function of $\sigma$, given by $\Omega_i
(\sigma ) = \bar \Omega + \sigma r_i$. Here, $r_i$ is a random
number extracted from a Gaussian distribution of zero mean and
unitary variance, chosen once for each oscillator and all values of
$\sigma$. In this way, individual natural frequencies maintain their
relative difference with $\bar \Omega$ as $\sigma$ varies. We take,
as above, $\bar \Omega=1$.

\begin{figure}
\includegraphics[width=\columnwidth]{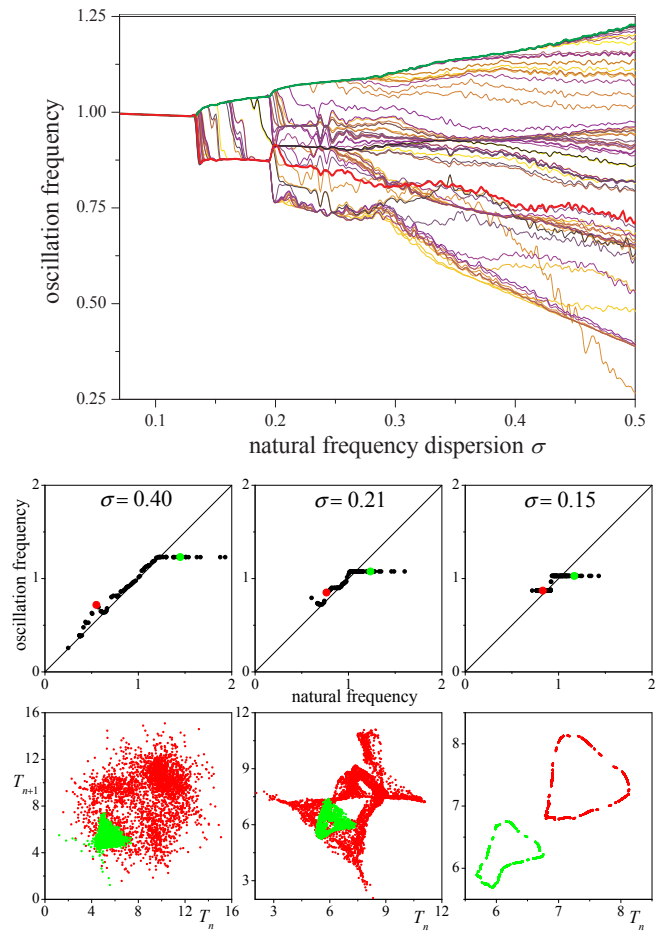}
\caption{(Color online) Upper panel: Individual oscillation
frequencies in an ensemble of $N=100$ oscillators, as functions of
the mean square dispersion of natural frequencies $\sigma$. The two
heavier lines (green/light grey and red/dark grey) correspond to two
oscillators selected for more detailed analysis (see text). Middle
panels: Oscillation frequencies as functions of natural frequencies
for all oscillators in the ensemble, and three values of $\sigma$.
The two selected oscillators are marked with their respective
colors/shades. Lower panels: Recurrence plots for the time intervals
$T_n$ between maxima in the coordinates of the two selected
oscillators, for the same values of $\sigma$ as in the middle
panels.} \label{fig9}
\end{figure}

The upper panel of Fig.~\ref{fig9} shows the oscillation frequency
of individual oscillators, calculated from the average interval
between their coordinate's maxima, as a function of the natural
frequency dispersion $\sigma$. Each curve corresponds to an
oscillator; heavier curves (green/light grey and red/dark grey)
stand for the frequencies of two oscillators whose individual
dynamics are analyzed in more detail below. Oscillation frequencies
have been evaluated at $2000$ values of $\sigma$ within the limits
of the plot, and then interpolated by means of spline functions in
order to smooth out some sharp fluctuations, especially in the
large-$\sigma$ range.

The plot shows how, as the dispersion $\sigma$ decreases,
increasingly many oscillators aggregate into clusters. Close
inspection of individual curves reveals the considerable complexity
of the process. While many oscillators join their nearest clusters,
others perform substantial frequency excursions before becoming
integrated into a cluster --see, for instance, the oscillator with
the lowest oscillation frequency for $\sigma=0.5$. Some small groups
of oscillators with neighboring frequencies seemingly coalesce into
well defined clusters, only to become dispersed as $\sigma$
decreases further, probably due to the disrupting effect of larger
and more coherent groups. Once the ensemble is split into just a few
clusters, their mutual collapse occurs within narrow intervals of
the dispersion, as seen to happen for $\sigma \approx 0.2$. For
$0.13 \lesssim \sigma \lesssim 0.2$, in turn, the migration of a few
oscillators from one cluster to another is especially noticeable,
though this phenomenon is also observed in other zones of the plot.
The final collapse into a single cluster takes place at $\sigma
\approx 0.13$, where --as we have found above, from the results
displayed in Fig.~\ref{fig8}-- the order parameter attains the
theoretical value corresponding to full synchronization.

The development of clusters is apparent in the plots of the middle
panels of Fig.~\ref{fig9}, where individual oscillation frequencies
are plotted versus the respective natural frequencies for three
selected values of the dispersion of natural frequencies. In this
kind of plot, where each dot represents a single oscillator,
plateaus of constant oscillation frequencies stand for clusters
formed by mutually synchronized oscillators. On the other hand, dots
near the diagonal correspond to oscillators which remain almost
immune to the collective force exerted by the ensemble. For
$\sigma=0.40$ (left panel), most of the population is
unsynchronized, although a well developed cluster is already present
at large frequencies. This cluster contains $28$ of the $100$
oscillators, whose oscillation frequencies differ by about one part
in $10^4$. There are, however, at least three more groups with
similarly close frequencies, but none of them contains more than
four oscillators. For $\sigma=0.21$ (central panel), the
large-frequency cluster has grown to encompass $42$ oscillators,
while a cluster with $13$ oscillators has appeared at intermediate
frequencies. A few smaller groups, of about five synchronized
oscillators each, are also present. For $\sigma=0.15$ (right panel),
all but three oscillators are distributed among two clusters. Of
these three oscillators --which, as we can see from the upper panel,
are migrating between the clusters as $\sigma$ varies-- two are
mutually synchronized, with oscillation frequencies differing by one
part in $10^5$. The large, low and high-frequency clusters contain
$32$ and $65$ oscillators, respectively.

Finally, in order to have a glance at the dynamics of individual
oscillators, we have constructed recurrence plots for the time
intervals $T_n$ between successive maxima in their coordinates --the
same intervals used to numerically evaluate the oscillation
frequencies. We have selected two oscillators, which we denote by A
and B, with natural frequencies at both sides of the mean value
$\bar \Omega=1$: $\Omega_{\rm A} \approx 1.17$ and $\Omega_{\rm B}
\approx 0.83$. In the plots of Fig.~\ref{fig9}, they are represented
by two colors/shades: green/light grey and red/dark grey,
respectively.

The lower panels in Fig.~\ref{fig9} display, for both oscillators,
the recurrence plots $T_{n+1}$ vs. $T_n$ along a few thousand
successive values of $n$ and for the same values of $\sigma$ as in
the middle panels. Note that, in this kind of plot, harmonic motion
would be represented by a single fixed point. For $\sigma=0.40$, the
irregular distribution of dots --with just a weak hint at an
underlying structure, given by the partially overlapping clouds--
suggests the occurrence of high-dimensional chaotic dynamics. We
point out that chaotic motion on a high-dimensional attractor is the
generic behavior expected for the ensemble when oscillations are not
fully synchronized. Our dynamical system, in fact, is dissipative
and nonlinear, and evolves in a phase space of many dimensions.
Clearly, this irregular motion is driven by the complicated
time-dependence of the coupling force, which integrates the
contributions coming from the mutually incoherent evolution of most
oscillators. Remarkably, while this force is the same for all
oscillators, the recurrence plot for $\sigma=0.40$ shows that its
effect on their individual dynamics can be quantitatively different.
Oscillator A, which already belongs to a large synchronized cluster
(see middle panels of Fig.~\ref{fig9}), is restricted to a
relatively small zone of the plot, fluctuating between $T_n\approx
4$ and $8$. Oscillator B, on the other hand, is not synchronized for
this value of $\sigma$, and the corresponding time intervals broadly
vary between $T_n \approx 2$ and $14$.

Oscillators A and B remain, respectively, synchronized and
unsynchronized for $\sigma=0.21$. The distribution of dots in the
recurrence plot is still irregular, but has become considerably more
structured than for higher $\sigma$. For oscillator B, in
particular, the picture is reminiscent of the attractor of a
low-dimensional chaotic map.  This hints at a considerable decrease
in the effective dimensionality of our system, as the organization
of the ensemble into synchronized clusters progresses.

For $\sigma=0.15$, the situation is qualitatively different. As
discussed above, practically all the ensemble is now divided into
just two synchronized clusters, comprising about one and two thirds
of the whole population. The coupling force is thus expected to
consist of essentially two components, each of them with the
frequency of one of the clusters. Accordingly, the distribution of
dots in the corresponding recurrence plot over well-defined closed
curves suggests quasiperiodic motion. Moreover, they are bounded to
a much smaller region of the plane, with variations in $T_n$ of
around one time unit. The collective organization in this partially
synchronized state has therefore led to a drastic decrease in the
dynamical complexity, with a large reduction in the dimensionality
of individual motion and a strong limitation to its deviation from
harmonic oscillations.

It should be clear that the above quantitative details about the
clustering dynamics of the oscillator ensemble --such as the number
of clusters at each value of $\sigma$, the number of oscillators in
each cluster, or the values of $\sigma$ at which the major cluster
collapses take place-- are specific to the ensemble obtained from a
particular realization of the random numbers $r_i$, which define the
individual natural frequencies. However, the generic picture arising
from the study of this particular case illustrates the features
typically expected for our system in the clustering regime.

\section{Discussion and conclusion}

In this paper, we have studied the synchronization dynamics of a
linear mechanical oscillator whose motion is sustained by a feedback
force, constructed through the conditioning of a signal produced by
the oscillator itself. This self-sustaining force --which, in an
experimental setup, can be generated electronically-- is an
amplitude-controlled, phase-shifted copy of the oscillator's
displacement from equilibrium. If the phase shift is appropriately
chosen, namely, if the force and the oscillator's velocity are
in-phase, the response of the system is optimized and the
oscillations attain maximal amplitude. Under the action of the
self-sustaining force, and for asymptotically long times, the
oscillator performs harmonic motion with a frequency determined by
the interplay between its mechanical parameters (mass, elastic
constant, damping coefficient) and those of the force itself.

Besides their plausible usefulness in the design of micromechanical
devices, on which we commented in the Introduction, oscillators of
this kind constitute an interesting class of non-standard mechanical
systems. In fact, the variables in their equation of motion are not
just the coordinate and the velocity, but also the oscillation
phase, whose instantaneous value cannot be unambiguously defined in
terms of the former. The dynamical properties of these oscillators
--in particular, those related to synchronized motion, which are
relevant to potential applications-- are therefore worth analyzing.

As a first step, we have considered the effect of an external
harmonic force on a single self-sustained oscillator. It is well
known that an ordinary linear mechanical oscillator responds to
harmonic forcing by asymptotically oscillating at exactly the same
frequency as the force, irrespectively of the difference with its
natural frequency. Our self-sustained oscillator, on the other hand,
synchronizes to the external force only if the difference between
the two frequencies is below a certain threshold. This
synchronization range grows as the oscillator's damping coefficient
and the amplitude of the external force increase. In this sense, the
self-sustained oscillator belongs to the wider class of oscillating
systems which can be entrained by external forces only if the
natural frequency and the forcing frequency are not too dissimilar.
However, in contrast to simpler systems --for instance, Kuramoto's
phase oscillators \cite{Kuramoto,synchro2}-- the frequency range
where synchronized dynamics is a solution to the equations of motion
does not coincide with the range where the same solution is stable.
Specifically, the existence range is always wider than the stability
range. When the amplitude of the external force is larger than that
of the self-sustaining force, in particular, synchronized solutions
exist for any frequency difference, while they are stable for
sufficiently small differences only.

We analyzed mutual synchronization of coupled oscillators, first, in
the case of two oscillators whose individual self-sustaining forces
were replaced by a common feedback force. This coupling signal
consisted of a linear combination of the two self-sustaining forces,
with equal weights for both contributions. The respective phase
shifts were also identical. Under these conditions, synchronized
solutions exist for arbitrary values of the oscillators'
self-sustained frequencies. The two synchronized oscillators perform
harmonic motion with a common frequency, whose value lies between
the two self-sustained frequencies. As in the case of
synchronization with an external force, however, synchronized motion
is not always stable. Mutual entrainment requires that the
self-sustained frequencies are sufficiently close to each other and,
again, the synchronization range grows with the oscillator's damping
coefficients. The requirement of small frequency differences to
ensure synchronization was stated several years ago for other
coupled systems whose individual dynamics exhibit, as in our case,
limit-cycle oscillations \cite{lc1,lc2,lc3}. It is however
interesting to point out that, in those systems, the synchronization
range is flanked, for sufficiently intense coupling, by a regime of
``oscillation death.'' In this regime, due to the effect of
coupling, the unstable fixed points inside individual limit cycles
become stable through an inverse Hopf bifurcation, and all
trajectories are asymptotically attracted toward them. The long-time
joint evolution is therefore trivial. In contrast, our
self-sustained oscillators do not possess fixed points in their
individual dynamics, and oscillation death is not observed.

For ensembles formed by many oscillators, the common coupling force
was constructed by combining equally weighted contributions from all
the elements. Coupling was therefore homogeneous and global over the
ensemble. Being a function of the individual oscillation phases, the
coupling force is straightforwardly related to Kuramoto's order
parameter for synchronization in populations of phase oscillators
\cite{Kuramoto}. This same order parameter can thus be used to
quantify the degree of coherence in our system. Assuming that the
ensemble is fully synchronized, with all oscillators moving with the
same frequency but shifted in phase from each other, it is possible
to show that the order parameter decreases as the dispersion in the
self-sustained frequencies over the ensemble grows, which reveals an
increasing dispersion in their relative phase shifts. In contrast
with the case of Kuramoto's phase oscillators, however, the order
parameter does not vanish for any finite value of the frequency
dispersion. This suggests that a synchronization transition similar
to that occurring in ensembles of phase oscillators is absent in our
system. Numerical resolution of the equations of motion, however,
shows that the state of full synchronization is not observed when
the frequency dispersion is large. Instead, the ensemble splits into
several groups of mutually synchronized oscillators, whose number
and size vary with the frequency dispersion, and with further
decrease of the order parameter. This is a typical regime of
clustering, reminiscent of the behavior of coupled chaotic dynamical
systems just below the synchronization transition \cite{synchro2},
or of ensembles of coupled oscillators with highly heterogeneous
frequency distribution \cite{penas}. In our system, this highly
degenerate collective state seems to be triggered by fluctuations in
the distribution of frequencies, with clusters forming where
frequencies accumulate randomly. The ensemble of interacting
self-sustained oscillators, in any case, shares with other coupled
systems the characteristic dynamical complexity of the clustering
regime.

In order to focus on the effect of the self-sustaining mechanism, we
have disregarded any other non-linear contribution to the individual
dynamics of the oscillators. Non-elastic forces, however, play a key
role in the functioning of self-sustained oscillators at microscopic
scales. As we have discussed in the Introduction, such forces are
responsible for the amplitude dependence of the oscillation
frequency --an undesired effect in frequency-control devices
\cite{af,af2,NatCom}. Hence, the next step in the study of this kind
of mechanical systems is to analyze the interplay between
self-sustaining and other non-linear forces, characterizing their
joint influence on synchronized motion. This is the subject of work
in progress.

\begin{acknowledgments}
We acknowledge financial support from CONICET (PIP 112-200801-76)
and ANPCyT (PICT 2011-0545), Argentina, and enlightening discussions
with Dar\'{\i}o Antonio, Hern\'an Pastoriza, and Daniel L\'opez.
\end{acknowledgments}

\bibliography{ea10954rev}

\begin{thebibliography}{24}
\expandafter\ifx\csname natexlab\endcsname\relax\def\natexlab#1{#1}\fi
\expandafter\ifx\csname bibnamefont\endcsname\relax
  \def\bibnamefont#1{#1}\fi
\expandafter\ifx\csname bibfnamefont\endcsname\relax
  \def\bibfnamefont#1{#1}\fi
\expandafter\ifx\csname citenamefont\endcsname\relax
  \def\citenamefont#1{#1}\fi
\expandafter\ifx\csname url\endcsname\relax
  \def\url#1{\texttt{#1}}\fi
\expandafter\ifx\csname urlprefix\endcsname\relax\def\urlprefix{URL }\fi
\providecommand{\bibinfo}[2]{#2}
\providecommand{\eprint}[2][]{\url{#2}}

\bibitem[{\citenamefont{Craighead}(2000)}]{sci}
\bibinfo{author}{\bibfnamefont{H.~G.} \bibnamefont{Craighead}},
  \bibinfo{journal}{Science} \textbf{\bibinfo{volume}{290}},
  \bibinfo{pages}{1532} (\bibinfo{year}{2000}).

\bibitem[{\citenamefont{Ekinci and Roukes}(2005)}]{rev}
\bibinfo{author}{\bibfnamefont{K.~L.} \bibnamefont{Ekinci}} \bibnamefont{and}
  \bibinfo{author}{\bibfnamefont{M.~L.} \bibnamefont{Roukes}},
  \bibinfo{journal}{Rev.\ Sci.\ Instrum.} \textbf{\bibinfo{volume}{76}},
  \bibinfo{pages}{061101} (\bibinfo{year}{2005}).

\bibitem[{\citenamefont{Greywall et~al.}(1994)\citenamefont{Greywall, Yurke,
  Busch, Pargellis, and Willett}}]{Greywall}
\bibinfo{author}{\bibfnamefont{D.~S.} \bibnamefont{Greywall}},
  \bibinfo{author}{\bibfnamefont{B.}~\bibnamefont{Yurke}},
  \bibinfo{author}{\bibfnamefont{P.~A.} \bibnamefont{Busch}},
  \bibinfo{author}{\bibfnamefont{A.~N.} \bibnamefont{Pargellis}},
  \bibnamefont{and} \bibinfo{author}{\bibfnamefont{R.~L.}
  \bibnamefont{Willett}}, \bibinfo{journal}{Phys.\ Rev.\ Lett.}
  \textbf{\bibinfo{volume}{72}}, \bibinfo{pages}{2992} (\bibinfo{year}{1994}).

\bibitem[{\citenamefont{Yurke et~al.}(1995)\citenamefont{Yurke, Greywall,
  Pargellis, and Busch}}]{Yurke}
\bibinfo{author}{\bibfnamefont{B.}~\bibnamefont{Yurke}},
  \bibinfo{author}{\bibfnamefont{D.~S.} \bibnamefont{Greywall}},
  \bibinfo{author}{\bibfnamefont{A.~N.} \bibnamefont{Pargellis}},
  \bibnamefont{and} \bibinfo{author}{\bibfnamefont{P.~A.} \bibnamefont{Busch}},
  \bibinfo{journal}{Phys.\ Rev.} \textbf{\bibinfo{volume}{51}},
  \bibinfo{pages}{4211} (\bibinfo{year}{1995}).

\bibitem[{\citenamefont{Jos\'e and Saletan}(1998)}]{reso}
\bibinfo{author}{\bibfnamefont{J.~V.} \bibnamefont{Jos\'e}} \bibnamefont{and}
  \bibinfo{author}{\bibfnamefont{E.~J.} \bibnamefont{Saletan}},
  \emph{\bibinfo{title}{Classical Dynamics: A Contemporary Approach}}
  (\bibinfo{publisher}{Cambridge University Press},
  \bibinfo{address}{Cambridge}, \bibinfo{year}{1998}).

\bibitem[{\citenamefont{Antonio et~al.}(2012)\citenamefont{Antonio, Zanette,
  and L\'opez}}]{NatCom}
\bibinfo{author}{\bibfnamefont{D.}~\bibnamefont{Antonio}},
  \bibinfo{author}{\bibfnamefont{D.~H.} \bibnamefont{Zanette}},
  \bibnamefont{and} \bibinfo{author}{\bibfnamefont{D.}~\bibnamefont{L\'opez}},
  \bibinfo{journal}{Nat.\ Commun.} \textbf{\bibinfo{volume}{3}},
  \bibinfo{pages}{802} (\bibinfo{year}{2012}).

\bibitem[{\citenamefont{Ward and Duwel}(2011)}]{noise}
\bibinfo{author}{\bibfnamefont{P.}~\bibnamefont{Ward}} \bibnamefont{and}
  \bibinfo{author}{\bibfnamefont{A.}~\bibnamefont{Duwel}},
  \bibinfo{journal}{IEEE Trans.\ Ultrason.\ Ferroelectr.\ Freq.\ Control}
  \textbf{\bibinfo{volume}{58}}, \bibinfo{pages}{195} (\bibinfo{year}{2011}).

\bibitem[{\citenamefont{Agarwal et~al.}(2007)\citenamefont{Agarwal, Mehta,
  Candler, Chandorkar, Kim, Hopcroft, Melamud, Bahl, Yama, Kenny et~al.}}]{af}
\bibinfo{author}{\bibfnamefont{M.}~\bibnamefont{Agarwal}},
  \bibinfo{author}{\bibfnamefont{H.}~\bibnamefont{Mehta}},
  \bibinfo{author}{\bibfnamefont{R.~N.} \bibnamefont{Candler}},
  \bibinfo{author}{\bibfnamefont{S.}~\bibnamefont{Chandorkar}},
  \bibinfo{author}{\bibfnamefont{B.}~\bibnamefont{Kim}},
  \bibinfo{author}{\bibfnamefont{M.~A.} \bibnamefont{Hopcroft}},
  \bibinfo{author}{\bibfnamefont{R.}~\bibnamefont{Melamud}},
  \bibinfo{author}{\bibfnamefont{G.}~\bibnamefont{Bahl}},
  \bibinfo{author}{\bibfnamefont{G.}~\bibnamefont{Yama}},
  \bibinfo{author}{\bibfnamefont{T.~W.} \bibnamefont{Kenny}},
  \bibnamefont{et~al.}, \bibinfo{journal}{J.\ Appl.\ Phys.}
  \textbf{\bibinfo{volume}{102}}, \bibinfo{pages}{074903}
  (\bibinfo{year}{2007}).

\bibitem[{\citenamefont{Agarwal et~al.}(2008)\citenamefont{Agarwal, Chandorkar,
  Mehta, Candler, Ki, and Hopcroft}}]{af2}
\bibinfo{author}{\bibfnamefont{M.}~\bibnamefont{Agarwal}},
  \bibinfo{author}{\bibfnamefont{S.~A.} \bibnamefont{Chandorkar}},
  \bibinfo{author}{\bibfnamefont{H.}~\bibnamefont{Mehta}},
  \bibinfo{author}{\bibfnamefont{R.~N.} \bibnamefont{Candler}},
  \bibinfo{author}{\bibfnamefont{B.}~\bibnamefont{Ki}}, \bibnamefont{and}
  \bibinfo{author}{\bibfnamefont{M.~A.} \bibnamefont{Hopcroft}},
  \bibinfo{journal}{Appl.\ Phys.\ Lett.} \textbf{\bibinfo{volume}{92}},
  \bibinfo{pages}{104106} (\bibinfo{year}{2008}).

\bibitem[{\citenamefont{Jos\'e and Saletan}(2007)}]{Talman}
\bibinfo{author}{\bibfnamefont{J.~V.} \bibnamefont{Jos\'e}} \bibnamefont{and}
  \bibinfo{author}{\bibfnamefont{E.~J.} \bibnamefont{Saletan}},
  \emph{\bibinfo{title}{Geometric Mechanics. Toward a Unification of Classical
  Physics}} (\bibinfo{publisher}{Wiley-VHC}, \bibinfo{address}{Weinheim},
  \bibinfo{year}{2007}).

\bibitem[{\citenamefont{Benedetto}(1996)}]{Hilbert}
\bibinfo{author}{\bibfnamefont{J.~J.} \bibnamefont{Benedetto}},
  \emph{\bibinfo{title}{Harmonic Analysis and Applications}}
  (\bibinfo{publisher}{CRC Press}, \bibinfo{address}{Boca Raton},
  \bibinfo{year}{1996}).

\bibitem[{\citenamefont{Pikovsky et~al.}(2003)\citenamefont{Pikovsky,
  Rosemblum, and Kurths}}]{synchro1}
\bibinfo{author}{\bibfnamefont{A.}~\bibnamefont{Pikovsky}},
  \bibinfo{author}{\bibfnamefont{M.}~\bibnamefont{Rosemblum}},
  \bibnamefont{and} \bibinfo{author}{\bibfnamefont{J.}~\bibnamefont{Kurths}},
  \emph{\bibinfo{title}{Synchronization: A Universal Concept in Nonlinear
  Sciences}} (\bibinfo{publisher}{Cambridge University Press},
  \bibinfo{address}{Cambridge}, \bibinfo{year}{2003}).

\bibitem[{\citenamefont{Manrubia et~al.}(2004)\citenamefont{Manrubia,
  Mikhailov, and Zanette}}]{synchro2}
\bibinfo{author}{\bibfnamefont{S.~C.} \bibnamefont{Manrubia}},
  \bibinfo{author}{\bibfnamefont{A.~S.} \bibnamefont{Mikhailov}},
  \bibnamefont{and} \bibinfo{author}{\bibfnamefont{D.~H.}
  \bibnamefont{Zanette}}, \emph{\bibinfo{title}{Emergence of Dynamical Order.
  Synchronization Phenomena in Complex Systems}} (\bibinfo{publisher}{World
  Scientific}, \bibinfo{address}{Singapore}, \bibinfo{year}{2004}).

\bibitem[{\citenamefont{Bittanti and Colaneri}(2010)}]{Floquet}
\bibinfo{author}{\bibfnamefont{S.}~\bibnamefont{Bittanti}} \bibnamefont{and}
  \bibinfo{author}{\bibfnamefont{P.}~\bibnamefont{Colaneri}},
  \emph{\bibinfo{title}{Periodic Systems: Filtering and Control}}
  (\bibinfo{publisher}{Springer}, \bibinfo{address}{Berlin},
  \bibinfo{year}{2010}).

\bibitem[{\citenamefont{Kuramoto}(2003)}]{Kuramoto}
\bibinfo{author}{\bibfnamefont{Y.}~\bibnamefont{Kuramoto}},
  \emph{\bibinfo{title}{Chemical Oscillations, Waves, and Turbulence}}
  (\bibinfo{publisher}{Courier Dover}, \bibinfo{address}{Reading},
  \bibinfo{year}{2003}).

\bibitem[{\citenamefont{Kaneko}(1990)}]{clus1a}
\bibinfo{author}{\bibfnamefont{K.}~\bibnamefont{Kaneko}},
  \bibinfo{journal}{Physica D} \textbf{\bibinfo{volume}{41}},
  \bibinfo{pages}{137} (\bibinfo{year}{1990}).

\bibitem[{\citenamefont{Kaneko}(1994)}]{clus1b}
\bibinfo{author}{\bibfnamefont{K.}~\bibnamefont{Kaneko}},
  \bibinfo{journal}{Physica D} \textbf{\bibinfo{volume}{75}},
  \bibinfo{pages}{55} (\bibinfo{year}{1994}).

\bibitem[{\citenamefont{Zanette and Mikhailov}(1998{\natexlab{a}})}]{clus2a}
\bibinfo{author}{\bibfnamefont{D.~H.} \bibnamefont{Zanette}} \bibnamefont{and}
  \bibinfo{author}{\bibfnamefont{A.~S.} \bibnamefont{Mikhailov}},
  \bibinfo{journal}{Phys.\ Rev.\ E} \textbf{\bibinfo{volume}{57}},
  \bibinfo{pages}{276} (\bibinfo{year}{1998}{\natexlab{a}}).

\bibitem[{\citenamefont{Zanette and Mikhailov}(1998{\natexlab{b}})}]{clus2b}
\bibinfo{author}{\bibfnamefont{D.~H.} \bibnamefont{Zanette}} \bibnamefont{and}
  \bibinfo{author}{\bibfnamefont{A.~S.} \bibnamefont{Mikhailov}},
  \bibinfo{journal}{Phys.\ Rev.\ E} \textbf{\bibinfo{volume}{58}},
  \bibinfo{pages}{872} (\bibinfo{year}{1998}{\natexlab{b}}).

\bibitem[{\citenamefont{Zanette and Mikhailov}(2000)}]{clus2c}
\bibinfo{author}{\bibfnamefont{D.~H.} \bibnamefont{Zanette}} \bibnamefont{and}
  \bibinfo{author}{\bibfnamefont{A.~S.} \bibnamefont{Mikhailov}},
  \bibinfo{journal}{Phys.\ Rev.\ E} \textbf{\bibinfo{volume}{62}},
  \bibinfo{pages}{R7571} (\bibinfo{year}{2000}).

\bibitem[{\citenamefont{Aronson et~al.}(1990)\citenamefont{Aronson, Ermentrout,
  and Kopell}}]{lc1}
\bibinfo{author}{\bibfnamefont{D.~G.} \bibnamefont{Aronson}},
  \bibinfo{author}{\bibfnamefont{G.~B.} \bibnamefont{Ermentrout}},
  \bibnamefont{and} \bibinfo{author}{\bibfnamefont{N.}~\bibnamefont{Kopell}},
  \bibinfo{journal}{Physica D} \textbf{\bibinfo{volume}{41}},
  \bibinfo{pages}{403} (\bibinfo{year}{1990}).

\bibitem[{\citenamefont{Matthews and Strogatz}(1990)}]{lc2}
\bibinfo{author}{\bibfnamefont{P.~C.} \bibnamefont{Matthews}} \bibnamefont{and}
  \bibinfo{author}{\bibfnamefont{S.~H.} \bibnamefont{Strogatz}},
  \bibinfo{journal}{Phys.\ Rev.\ Lett.} \textbf{\bibinfo{volume}{65}},
  \bibinfo{pages}{1701} (\bibinfo{year}{1990}).

\bibitem[{\citenamefont{Reddy et~al.}(1998)\citenamefont{Reddy, Sen, and
  Johnston}}]{lc3}
\bibinfo{author}{\bibfnamefont{D.~V.~R.} \bibnamefont{Reddy}},
  \bibinfo{author}{\bibfnamefont{A.}~\bibnamefont{Sen}}, \bibnamefont{and}
  \bibinfo{author}{\bibfnamefont{G.~L.} \bibnamefont{Johnston}},
  \bibinfo{journal}{Phys.\ Rev.\ Lett.} \textbf{\bibinfo{volume}{80}},
  \bibinfo{pages}{5109} (\bibinfo{year}{1998}).

\bibitem[{\citenamefont{Mikhailov et~al.}(2004)\citenamefont{Mikhailov,
  Zanette, Zhai, Kiss, and Hudson}}]{penas}
\bibinfo{author}{\bibfnamefont{A.~S.} \bibnamefont{Mikhailov}},
  \bibinfo{author}{\bibfnamefont{D.~H.} \bibnamefont{Zanette}},
  \bibinfo{author}{\bibfnamefont{Y.~M.} \bibnamefont{Zhai}},
  \bibinfo{author}{\bibfnamefont{I.~Z.} \bibnamefont{Kiss}}, \bibnamefont{and}
  \bibinfo{author}{\bibfnamefont{J.~L.} \bibnamefont{Hudson}},
  \bibinfo{journal}{Proc.\ Nat.\ Acad.\ Sci.\ USA}
  \textbf{\bibinfo{volume}{101}}, \bibinfo{pages}{10890}
  (\bibinfo{year}{2004}).

\end{thebibliography}

\end{document}